\documentclass[12pt]{article}
\usepackage{graphicx}
\usepackage{amssymb,amsmath,amsfonts,palatino,amsthm}
\usepackage{amssymb}
\usepackage{epstopdf}
\newcommand{\f}{\begin{equation}}
\newcommand{\ff}{\end{equation}}

\begin{document}

\title{Unification of the state with the dynamical law \\}
\author{Lee Smolin\thanks{lsmolin@perimeterinstitute.ca} 
\\
\\
Perimeter Institute for Theoretical Physics,\\
31 Caroline Street North, Waterloo, Ontario N2J 2Y5, Canada}
\date{\today}
\maketitle

\begin{abstract}
We address the question of  why particular laws were selected for the universe,  by proposing a mechanism for laws to evolve.  Normally in physical theories, timeless laws act on time-evolving states. We propose that this is an approximation, good on time scales shorter than cosmological scales, beyond which laws and states are merged into a single entity that evolves in time.  
Furthermore the approximate distinction between laws and states, when it does emerge, is dependent on the initial conditions. These ideas are illustrated in a simple matrix model.


\end{abstract}

\newpage

\tableofcontents

\section{Introduction}

Physics has for most of its history been primarily concerned with finding out what the laws of nature are.  While we still do not have a completely unified theory of physics, our understanding of the laws of nature has advanced to the point where we are not only interested in what the laws are, but why these are the laws, and not others.
This problem has become urgent since the discovery of the landscape of string theories\cite{andy,LOTC}.  The hope that a theory that unifies gravity and the standard model of particle physics would be unique, in a way that leads to unique predictions for beyond the standard model physics,  seems difficult to sustain in the face of a vast or infinite number of apparently equally consistent string vacua.  Even if one is not confident that string theory is the right framework for unification, no framework has appeared which would answer the why these laws question.   

The realization that we would sooner or later have to explain how and why the laws we observe governing our universe were chosen is not new.  The issue was emphasized by John Wheeler\cite{JAW}, but the concern is much older and goes back to Leibniz's Principle of Sufficient Reason.   As the American pragmatist philosopher Charles Sanders Pierce wrote in 1893, nothing is so needing of rational explanation than laws of nature.  Pierce goes on to say\cite{Pierce},

\begin{quotation}

{\it To suppose universal laws of nature capable of being apprehended by the mind and yet having no reason for their special forms, but standing inexplicable and irrational, is hardly a justifiable position. Uniformities are precisely the sort of facts that need to be accounted for. Law is par excellence the thing that wants a reason. Now the only possible way of accounting for the laws of nature, and for uniformity in general, is to suppose them results of evolution.} 

\end{quotation}

In contemporary work, all the present attempts to understand how laws may have been chosen from a landscape of possible laws evoke, in one way or another, the notion that the effective low energy laws change on cosmological time scales.  This includes  eternal inflation\cite{EI} and cosmological natural selection\cite{CNS}.  It is notable that both require a notion of time to give meaning to the evolution of effective laws.  In cosmological natural selection a time is required to count generations and give sense to an ensemble of universes on the landscape at a fixed time, in eternal inflation it appears necessary to impose a measure which is related to a notion of time on the multiverse.  

The view that laws have to evolve in a physically real, non-emergent time, in order to have a scientific explanation of why these laws has been developed in work in progress with Roberto Mangabeira Unger\cite{robertolee}.
This note explores a suggestion made there,  which is that the evolution of laws implies a breakdown of the distinction between law and state.  Another way to say this is that there is an enlarged notion of state-a metastate- which codes information needed to specify both an effective law and an effective state, that the effective law acts on.
The whole metastate evolves in time, and the distinction between effective law and effective state can only be made for certain time scales. How long these time scales are, as well as the effective laws, are determined by the initial metastate.  The effective laws evolve with the state, but they  evolve slowly,
compared to other information captured by the state. 

Hence, on short time scales, and to a certain approximation, one can distinguish a slowly varying effective law which generates faster evolution of an approximate state.   On longer time scales the more precise picture is that there is a notion of a meta-state, which codes both the effective law and the effective state.  

The purpose of this paper is to present a simple matrix model where this idea is realized. But in realizing this idea we have to confront an issue that arises in any scenario in which laws of physics evolve. 
In both cosmological natural selection\cite{CNS} and eternal inflation\cite{EI} there is posited a dynamical mechanism whereby a population of regions of the universe with different laws evolves, giving rise to an evolving distribution on a landscape, or space of laws, $\cal L$  \cite{fitness}.  The evolution of laws on the landscape is then driven by a {\it metalaw.}  Even if not precisely specified, this metalaw  becomes a key part of the explanation of why these laws.  

In the case of cosmological natural selection, the metalaw is approximate and effective and involves small random changes in the parameters of the standard model.  This is analogous to an effective dynamics for evolution of phylogeny in biology. In the case of eternal inflation the metalaw is tunneling from false vacua with amplitudes given by string theory. 

However, these scenarios have a weak point. The postulation of metadynamics on the landscape is a scientific hypothesis.  How is it to be justified? If there is no principle which determines the law, it is not likely there will be a principle which determines the metalaw.  And how is the proposed metalaw to be tested?    One can easily imagine different hypotheses for the action of metalaws on the spaces of observable parameters.  How are these to be compared, when we see in our past at best one instance of the metalaws acting?  Someone may claim that the evolution on the landscape is driven by some fundamental version of string theory.  Someone else may claim the evolution on the landscape is fundamentally stochastic (and why not-so is quantum theory?) and driven only by a simple set of rules.  How are we to determine scientifically which is right?  

Worse, we may have to postulate some metalandscape of metalaws on which a meta-meta-law acts to  govern the choice of the metalaw. 
There is clearly a  danger of an infinite regress here.

On the other hand, if one does not specify a metalaw one explains nothing.  We call this the {\it metalaws dilemma\cite{robertolee}.}  

It is important to have a precise idea of what is going wrong when we encounter this kind of dilemma.  
Normally in physics we specify a theory in two steps.  First we specify the configuration or a phase space, $\cal C$, which is a timeless space of possible configurations a system may have at one time. 
Then we specify the laws of motion, which generates the possible lawful trajectories of the system on $\cal C$.  If we append to this the landscape of possible laws we have two timeless configuration spaces: that of configurations and that of laws. 

This formulation of laws of nature can be called the {\it Newtonian paradigm} because it is the basic framework of laws of motion introduced by Newton\cite{robertolee}.  The Newtonian paradigm is also the framework for modern quantum mechanics, quantum field theory, and general relativity.  In each case there is a timeless space of states acted on by a timeless law.  

The Newtonian paradigm is the proper setting for most of physics, which concerns small subsystems of the universe.  But when we attempt to scale it up to a description of the universe as a whole it leads to unanswerable questions such as why these laws and not others and what caused the initial conditions.  No theory  formulated within the Newtonian paradigm can answer these questions because it takes the laws and initial conditions as inputs.    When we attempt to invent a theory of evolution on a landscape of theories, but stay within the Newtonian paradigm, we end up with puzzles and paradoxes. 

Part of the problem is the following. The Newtonian paradigm is based on a strict separation of the roles of law and initial conditions.  This is justified by the fact that we can operationally distinguish the influence of the choice of laws from the choice of initial conditions, by doing experiments many times varying the initial conditions. 
Operationally, what we mean by a law is some feature of the evolution which is invariant or conserved when we vary the initial conditions.  So the experimental context that gives meaning to theories formulated in the Newtonian paradigm is the study of small subsystems of the universe, where we can repeat an experiment as many times as needed.

In cosmology there is only a single history, so we loose the ability to do an experiment over and over again, while varying the initial conditions. So we have no operational way to absolutely distinguish the influence of the choice of laws from the choice of initial conditions.  When we attempt to impose the Newtonian paradigm on the interpretation of cosmological data, and ask questions that assume a strict separation between the role of law and the role of initial conditions, we end up asking confused questions that have no clear answers.  

We call this running into the {\it cosmological fallacy}\cite{robertolee}, which is the mistake of extending a method that is designed to study small subsystems of the universe that come  in many copies to the universe as a whole.  To usefully apply a theory in the Newtonian paradigm to a system we 
require data from many repetitions of an experiment to give operational meaning to its basic terms, and in particular, to separate out the role of laws from initial conditions.  But in the cosmological case, the data does not allow that distinction to be made\footnote{The multiverse seems at first to be a way to avoid this, because it makes our universe one out of many and so appears to reproduce the context needed to make sense of the separation between law and initial condition.  This however, cannot succeed so long as we have no data about the other universes, because there is still no operational basis for the distinction between law and initial conditions.  The only exception, is  special cases where each universe in the ensemble shares a property-then one can check the theory by seeing if our universe has that property. This is the strategy of cosmological natural selection. It also leads to a single prediction for eternal inflation, which is that all universes in the mutliverse have $k=-1$.}. 

It is probably wiser to not impose a paradigm for dynamical law on the cosmological data that is based on a distinction that cannot be made within that data.  That is, once we loose the ability to distinguish the role of law and initial conditions in the data, because we have just one case in cosmology, we are probably going to make more progress if we search for a framework for physical theory that does not rely on the distinction between law and initial condition being absolute.  

What is then needed is a new paradigm for dynamics on a cosmological scale.  In this new framework, the absolute distinction between laws and states, or laws and initial conditions, which underlies the Newtonian paradigm can be transcended. That distinction will be seen to be an artifact of descriptions of small subsystems of the universe, and breaks down on cosmological time scales.   The challenge is to introduce such a framework without falling into a vicious circle or the metalaws dilemma.  The purpose of this paper is to explore one possible form that such a new approach to cosmological dynamics may take.  

In previous work\cite{universality}, I proposed a possible resolution to these conundra, which is that there could be a notion of universality of metalaws, analogous to universality in the theory of computation. The idea is that any metalaw which could serve as such is equivalent to any other.  Computation is universal because any computer can emulate any other exactly.  The proposal in \cite{universality} is that any metalaw worthy of that name can emulate any other, because they will lead to the same predictions for the evolution of laws.    In \cite{universality}, I made a proposal for such a universal metalaw in the context of a matrix model.  

In this note, I propose a model for another approach to the metalaws dilemma, which is that the distinction between states and laws breaks down. This new proposal is also realized in a simple matrix model. Instead of timeless law determining evolution on a timeless space of states, we have a single evolution which cannot be precisely broken down into law and state.  
Formally, what this means is to embed the configuration space of states, $\cal C$ and the landscape parameterizing laws, $\cal L$ into a single meta-configuration 
space, $\cal M$.   The distinction between law and state must then be both approximate and dependent on initial conditions.   

There is, it must be granted, an evolution rule on $\cal M$, but we can choose an evolution rule that is almost entirely fixed by some natural assumptions. The remaining freedom is, I conjecture, accounted for by the principle of universality, which I just described.    Because the complexity of the effective law is now coded into the state, the meta-law can be very simple, because all it has to do is to generate a sequence of matrices, in which the differences from one to the next are small.  The metalaw dillemma is addressed by showing that the form of this rule is almost completely fixed by some natural assumptions, with the remaining freedom plausibly accounted for by universality. 

In this model of a metatheory, the metastate is captured in a large matrix, $X$, which we take to be antisymmetric and valued in the integers.  It might describe a labeled graph.  The metalaw is a simple algorithm that yields a sequence of matrices, $X_n$.  The rule is that $X_n$ is gotten by adding to a linear combination of $X_{n-1}$ and
$X_{n-2}$ their commutator $[X_{n-1},X_{n-2}]$.  Given the first two matrixes, $X_0$ and $X_1$,  the sequence is determined.   This is more like a simple instruction in computer science than a law of physics, and we are able to argue it is almost unique, given a few simple conditions.  

That almost unique evolution rule acts on a configuration space of matrices, whose interpretation depends on a separation of time scales.  For certain initial configurations- there will be a long time scale, $T_{Newton}$ such that, for times shorter than $T_{Newton}$, the dynamics can be approximately described by a fixed law acting on a fixed space of states.  Both that law and that state are coded into the $X_n$.  But for longer times everything evolves, laws and states together, and it is impossible to cleanly separate what part of the evolution is changes in law and what part is changes in state.  Furthermore, which information in $\cal M$ evolves slowly, and goes into the specification of the approximate time independent law, and which evolves fast, and goes into the description of the time dependent state, is determined by the initial conditions.   

So the question of "why these laws" becomes subsumed into the question of "why these initial conditions" in a metatheory.  This does not yet solve the problem of explaining the particular features of the standard model and its parameters, but it gives a new methodology and strategy with which to search for the answer. 

Starting from the standard model, one might move in the direction of a metatheory be elevating all parameters to degrees of freedom.  This is something like what happens in the string landscape.  Here we make a simple model in which the meta-state is a large sparse matrix, perhaps representing the connections on a graph.  


In the next section we describe a simple  model which illustrates these ideas and show how it leads, for short time scales, to an approximate distinction between an effective law which governs the evolution of a state. 

\section{A minimal evolution rule}

We are interested in the most minimal evolution rule we can imagine which combines the theory and the state.  Let us specify the meta-state
by an $N \times N$ antisymmetric matrix of integers, $(X_n)_{ab}= - (X_n)_{ba}$.  We will consider the dimension $N$ to be large.  The $n$ refers to a succession of times, $n=0,1,2,...$ also labeled by integers.
 $(X_n)_{ab}$ might be taken to describe an adjacency matrix of a weighted, directed, graph, whose edges are labeled by integers.   This accords with the expectation that the fundamental variables in physics be relational.

 The idea is that there will be an evolution rule which specifies the series of matrices, given initial choices.
 The choice of this evolution rule  is fixed by the following ideas.  
 
 \begin{enumerate}
 
 \item{} The evolution rule should mimic second order differential equations, as these are basic to the dynamics of physical systems.  So two initial conditions should be required to generate the evolution.  We should then need to specify $X_0$ and $X_1$ to generate the sequence.  We are then interested in rules of the form.
 \f
 X_n = {\cal F}(X_{n-1}, X_{n-2})
 \label{r1}
 \ff
\item{}  The changes should be small from matrix to matrix, at least given suitable initial conditions.  This is needed so that there can be a long time scale on which some of the information in the matrixes are slowly varying.  This makes it possible to extract a notion of slowly varying law, acting on a faster varying state.  So we will ask that
 \f
 X = {\cal F}(X, X )
 \label{r2}
 \ff
 
\item{} We require that the evolution rule be non-linear, because non-linear laws are needed to code interactions in physics.  But we can always use the basic trick of matrix models of introducing auxiliary variables, by expanding the matrix, in order to lower the degree of non-linearity.  This accords with the fact that the field equations of general relativity and Yang-Mills theory can, by the use of auxilary variables, be expressed as 
quadratic equations\footnote{As for example in the Plebanski action.}\cite{universality}. The simplest non-linear evolution rule will then suffice, so we {\it  require a quadratic evolution rule.}  
 
\item{} Time reversal invariance, at least at the linear level.  
 
 \end{enumerate}
 
A simple evolution rule that realizes these is 
 \f
 \boxed{
 X_n = 2X_{n-1} - X_{n-2} +  [ X_{n-1}, X_{n-2}] }
 \label{e1}
 \ff
 This rule is not unique, but it is nearly so.  It is easy to derive the general rule satisfying the four requirements just mentioned.
 
The rule (\ref{r1}) can only have a linear term and a quadratic term.  The quadratic term must be a function of
 $X_{n-1}$ and $X_{n-2}$ that vanishes when they are equal and is antisymmetric.  The unique term that does this is the commutator  $[ X_{n-1}, X_{n-2}]$ .   When the commutator vanishes
there is only a linear term which by ( \ref{r2}) must be equal to a linear integral combination of $X_{n-1}$ and $X_{n-2}$.  The general evolution rule satisfying the first three requirements is then
\f
X_n =  a X_{n-1} + (1-a) X_{n-2} + g [ X_{n-1}, X_{n-2}] 
\label{gen-rule}
\ff
where $a$ and $g$ must be integers to keep the coefficients of $X_n$ integers.  

We pick $a=2$ and $g=1$ to get (\ref{e1}).  The justification for the choice of the linear term is time reversal invariance.  With this choice (\ref{e1}) can be written as
\f
\Delta^2 X_n \equiv  X_n +X_{n-2} -  2X_{n-1}  =  [ X_{n-1}, X_{n-2}] 
\ff
The linear term, $\Delta^2 X_n$ is invariant under a time reversal transformation around a time $n-1$, given by
\f
X_{n-1+\hat{a}} \leftrightarrow X_{n- 1 - \hat{a}}
\ff
under which $\Delta^2 X_n \rightarrow \Delta^2 X_n$

The whole dynamics is {\it approximately invariant}  under a related transformation $X_{n-1 +\hat{a}} \leftrightarrow -  X_{n- 1- \hat{a}}$. An exactly time invariant version of dynamics would be $ X_n = 2X_{n-1} - X_{n-2} +  [ X_{n}, X_{n-2}] $, but this is much harder to evolve.    

 
 Here is a way to understand how state and law are combined under this evolution rule.  Let us call the "Hamiltonian at time $n$", 
 \f
 H_n = X_{n-2}
 \ff
 and define the "state at time $n$" to be 
 \f
 \rho_n = X_n- X_{n-1}
 \label{rho}
 \ff
 We can define the rate of change of the state as 
 \f
 \Delta \rho_n = \rho_n - \rho_{n-1} = \Delta^2 X_n
 \label{ef}
 \ff
 Then the evolution rule (\ref{e1}) is expressed as
 \f
 \Delta \rho_n = [\rho_{n-1} , H_n ] 
 \label{ee2}
 \ff
 Thus it appears that the matrix we call $H_n$ is generating evolution on the state called $\rho$. 
 
 Another equivalent way to express the evolution is
 \f
 \Delta^2 \rho_n = [\Delta \rho_{n-1} , X_{n-2} ]
 \label{ee3}
 \ff 
 
 \subsection{Quasi-Hamiltonian evolution}

 Equations  (\ref{ef},\ref{ee2}) holds at all time steps. But this is not really Heisenberg evolution because the operator we are calling the Hamiltonian evolves
 as the state evolves.  But, as I will now show, if we choose the initial conditions so $\rho$ is in a certain sense small compared to $H$, then the
 $H$ evolves more slowly than $\rho$ and so for short times it appears as if the state is evolving with respect to a fixed Hamiltonian, so that 
  (\ref{rho},\ref{ee2})  are, for a finite time, well approximated by a Heisenberg-like equation of motion,
  \f
  \Delta \rho_n  = [ \rho_{n-1}, H_0 ].
  \label{hiesenberg}
  \ff
 
To show this we introduce a norm on matrices $||X||$ which is equal to the number of non-zero entries.  Then, if $p(X)$
is the probability that a matrix element is non-zero, then
\f
p(X) = \frac{||X||}{N(N-1)/2}
\ff

Pick an arbitrary time and call it $n=0$.  Call 
 \f
 X_0 = H_0,  \ \ \ \ \  \rho_1 = A
 \ff
 Then define
 \f
 \dot{A}= [A, H_0]  , \ddot{A} = [\dot{A}, H_0 ]  , \  \  \  \ . . . . A^{(p)}= [ A^{(p-1)}, H_0] 
 \ff

\subsection{The first steps}

Let us follow the first few steps of evolution
\begin{eqnarray}
& X_0 = H_0  & \rho_1 =A   \nonumber \\
& X_1 = H_0 +A  & \rho_2 =A+ \dot{A}   \nonumber \\
& X_2 = H_0 +2 A +\dot{A}  & \rho_3 =A+ 2\dot{A} +\ddot{A} + [\dot{A}, A] \nonumber \\
& X_3 = H_0 +3 A +3 \dot{A}  +\ddot{A} +  [\dot{A}, A]  & \rho_4 =A+ 3\dot{A} +3 \ddot{A} + A^{(3)}+  [\dot{A}, A] +  [\ddot{A}, A]  \nonumber \\
&& \ \ \ \ \ \ \ \ + [ 2\dot{A} +\ddot{A} +   [\dot{A}, A]   , 2A +\dot{A}  ] \nonumber  \\
&X_4 =  H_0 +4 A +6 \dot{A}  +4 \ddot{A} +  A^{(3)}+ 2 [\dot{A}, A]  
 & \nonumber \\
& \ \ \ \ \ \  +  [\ddot{A}, A]  + [ 2\dot{A} +\ddot{A} +   [\dot{A}, A]   , 2A +\dot{A}  ]  &
\end{eqnarray}

Clearly terms are rapidly proliferating.  To make sense of them, begin by noting that there are two kinds of terms in
the $\rho_n$s.  First there are terms that involve single powers of $A^{(p)}$.  These come from commutators with $H_0$ and can be considered
to be the effect of evolution with a fixed hamiltonian, $H_0$.  Then there are terms involving commutators of two or more $A^{(p)}$.  These register
the effect of the changing evolution law. As we will now show, there are natural choices of $H_0$ and $A$ such that the latter remain unimportant for
a large number of time steps.  

\subsection{Norms and probabilities}

 Let us pick $X_0 = H_0$  to be a random matrix chosen from the ensemble with 
 \f
 p(H_0) = \frac{1}{N}
 \ff
 so that it corresponds to the critical region in random graph theory of a graph which is minimally connected.  Then
 \f
 ||H_0||= N
 \ff
 We will pick $A$ to have a norm of order unity, so that $X_1$ differs from $X_0=H_0$ by just a few entries or links.  Then
 \f
  ||A||= M \in O(1) \ \ \ \mbox{so that} \ \   p(A) \approx \frac{M}{N^2}
 \ff
 Let us assume that there are no further correlations between $H_0$ and $A$ so that, 
 \f
 p(\dot{A}) = N p(A) p(H_0) \approx \frac{M}{N^2}
 \ff
 so that 
 \f
 ||\dot{A}|| \approx M
 \ff
 It then follows that all the 
 \f
  ||{A}^{(p)} || \approx M
 \ff 
Notice that because these ${A}^{(p)}$ are so sparse
\f
p ([ A, \dot{A} ] ) = p ([{A}^{(p)}, {A}^{(q)}]) = n p(A)^2 = \frac{M}{N^3}
\ff
Hence the norm of these commutators is 
\f
|| [{A}^{(p)}, {A}^{(q)}] || = \frac{M}{N} << 1
\ff
This means that there are no entries in most of these commutators.    

\subsection{Breakdown of the distinction between law and state}
 
Now after $n$ evolution steps we will have a time dependent Hamiltonian $H_n = X_{n-2}$ of the form
\f
H_n = H_0 + \delta H_n
\ff 
where the time dependent part has the form, 
\f
 \delta H_n =  \delta H_n(M) +  \delta H_n(M^2) + ...
 \label{deltaHn}
 \ff
 where $(M^p )$ signifies the terms of order $M^p$
The leading term collects terms of order $M$ which come from commutators of the form $[A^{(p)}, H_0]$.
 \f
 \delta H_n(M)  = \sum_{p=1}^{n-3} c_p A^{(p)} 
\ff
Here the $c_p$ are integer coefficients.  

Any one of the $A^{(p)}$ terms likely has no effect on the evolution of the $\rho$'s, but there are $n$ of them so $\delta H_n$ will start to be significant when
$n$ is large enough.  Hence, ignoring the $O(M^2)$ terms, we have
\f
p (  \delta H_n(M) ) = nN \frac{M}{N^2} \frac{1}{N} = \frac{nM}{N^2}
\ff
These will be negligible compared to the terms in $H_0$ if
\f
\frac{p (  \delta H_n(M) ) }{p (  H_0 ) } = \frac{nM}{N} <1
\ff
so the approximation in which we neglect the terms in $ \delta H_n(M)$ is good so long as 
\f
n< \frac{N}{M}.   
\label{n<N}
\ff
We can reach the same conclusion by computing the ratio of $p ( A^{(p)} )$ to  $p ( [A^{(p)},   A^{(q)} ] )$

Similarly we can compute the importance of the order $M^2$ terms in $\delta H_n$.  These come from commutators of the form $[{A}^{(p)}, {A}^{(q)}] $. Any one of these is  most likely vanishing, but there are $n^2$ of them.  

We have,
\f
p (  \delta H_n(M) ) = n^2 N \left ( \frac{M}{N^2} \right )^2 = \frac{n^2 M^2}{N^3}
\ff
These can be neglected relative to the entries in $H_0$ so long as 
\f
\frac{p (  \delta H_n(M^2 ) ) }{p (  H_0 ) } = \frac{n^2M^2}{N^2} <1
\ff
which leads us to the same condition (\ref{n<N}).  Indeed, the order $M^	q$ term in $\delta H_n$ comes from $q-1$ commutators of
factors $A^{(p)}$ for $p<q$, so these have probabilities
\f
p (  \delta H_n(M^q) ) = n^q N^{q-1} \left ( \frac{M}{N^2} \right )^q = \frac{n^q M^q}{N^{q+1}}
\ff
These are each negligible compared to the matrix elements of $H_0$ so long as (\ref{n<N}) holds.  One can also show this for the sum (\ref{deltaHn}).
Assuming the matrices are random so that the commutators are uncorrelated in the limit of large $N$ we can write,
\begin{eqnarray}
p ( \delta H_n) &=& p (  \delta H_n(M) ) + p ( \delta H_n(M^2) + ...  ) 
\nonumber \\
&=& \frac{1}{N} \left \{  \frac{nM}{N} +  \left (     \frac{nM}{N}        \right )^2    +....  \right \}
\nonumber \\
&<&  \frac{1}{N} \left \{  \sum_{q=1}^\infty  \left (     \frac{nM}{N}        \right )^q \right \} = 
 \frac{1}{N} \frac{ \frac{nM}{N}}{1- \frac{nM}{N}}
\end{eqnarray}
Hence 
\f
\frac{p (  \delta H_n ) }{p (  H_0 ) } <  \frac{ \frac{nM}{N}}{1- \frac{nM}{N}}
\ff
which is small so long as  (\ref{n<N}) holds.
This means that the Hamiltonian evolution law (\ref{hiesenberg}) is a good approximation to the exact dynamics so long as (\ref{n<N}) holds.

\section{Conclusions}

In the introduction of this paper we argued for that a cosmological theory must be formulated in a way in which the usual distinction between dynamics and state, or between kinematics and dynamics, breaks down\cite{robertolee}.  In the rest of this paper we illustrated these ideas with a simple toy model.  In it we addressed the problem of what determines the meta-law by which effective laws evolve by specifying four simple properties that almost completely determine it.  We conjecture that the remaining freedom is unimportant, because there may be a principle of universality among the remaining choices, in the sense that the predictions made by each of them can be mapped to each other.  

There remain of course open questions, among which are to demonstrate this conjecture of universality.

\section*{ACKNOWLEDGEMENTS}

This worked evolved from joint work with Roberto Mangabeira Unger\cite{robertolee} on the implications of the hypothesis that the laws of physics evolve, and I am very grateful to him for many suggestions and discussions as well as for critical comments on this manuscript.  I am grateful also to Andrew Albrecht, Andrzej Banbursk, Sabine Hossenfelder and Matthew Johnson for perceptive comments on the manuscript. I am also grateful to FQXi and the Templeton Foundation for generous support of my research.   Research at
Perimeter Institute for Theoretical Physics is supported in part by
the Government of Canada through NSERC and by the Province of
Ontario through MRI.


\begin{thebibliography}{99}

\bibitem{robertolee}Roberto Mangabeira Unger and Lee Smolin, book manuscript in preparation.  Some of the ideas in this project were published in L. Smolin,  {\it The unique universe: Against the timeless multiverse}, Physics World, June 2009, pps 21-26.

\bibitem{andy}Andrew Strominger, {\it Superstrings with torsion}, Nucl. Phys. B 274 (1986) 253-84. 

\bibitem{fitness}For the analogy to the fitness landscape of biology, see L. Smolin,   {\it Cosmology as a problem in critical phenomena}  
in the proceedings of the Guanajuato Conference
on Complex systems and binary networks, (Springer,1995),
eds. R. Lopez-Pena, R. Capovilla, R. Garcia-Pelayo,
H. Waalebroeck and F. Zertuche.
gr-qc/9505022; 

\bibitem{LOTC}Lee Smolin {\it Life of the Cosmos} Oxford University Press and Wiedenfeld and Nicolson, (1997).  

\bibitem{JAW}John Archibald Wheeler, in {\it Black holes, gravitational
waves and cosmology} eds. Martin Rees, Remo Ruffini and J. A. 
Wheeler, New York: Gordon and Breach, 1974.

\bibitem{Pierce}Charles Sanders Peirce,  {\it The architecture of theories}, The Monist, 1891, reprinted in {\it Philosophical Writings of Peirce}, ed. J. Buchler. New York: Dover, 1955.

\bibitem{EI}Alexander  Vilenkin (1983). {\it The birth of inflationary universes}. Phys. Rev. D27 (12): 2848. 

\bibitem{CNS} Lee Smolin,  {\it Did the universe evolve?}  Classical and Quantum Gravity 9 
(1992) 173-191.
     For recent updates, see Lee Smolin {\it The status of cosmological natural selection},
 hep-th/0612185, to appear in {\bf Beyond the Big Bang}, Springer Verlag, ed by Ruediger Vaas;  {\it Scientific alternatives to the anthropic principle,'},  arXiv:hep-th/0407213,
Contribution to "Universe or Multiverse", ed. by Bernard Carr et. al., published by Cambridge University Press; {\it Cosmological natural selection as the explanation for the complexity
of the universe}, Physica A: Statistical Mechanics and its Applications
Special issue: Complexity and Criticality: in memory of Per Bak (1947--2002) -
Edited by P. Alstrom, T. Bohr, K. Christensen, H. Flyvbjerg, M.H. Jensen, B.
Lautrup and K. Sneppen; {\it Using neutrons stars and primordial black holes to
test theories of quantum gravity}, astro-ph/9712189;  {\it On the fate of black hole singularities and the parameters
of the standard model}  submitted to Physical Review D.
 gr-qc/9404011.

\bibitem{universality}Lee Smolin, {\it Matrix universality of gauge and gravitational dynamics}, arXiv:0803.2926.         

\end{thebibliography}
\end{document}